\documentclass[pre,twocolumn,showpacs,preprintnumbers,superscriptaddress,amsmath,amssymb,floatfix,10pt]{revtex4-1}
\usepackage{graphics,epsfig}
\usepackage{amsmath,amssymb}
\usepackage{bm}
\usepackage{graphicx}
\usepackage[usenames]{color}
\usepackage{relsize}
\newcommand{\dif}{\mathrm{d}}%
\newcommand{\fdif}{\operatorname{\delta}}
\newcommand{\Fdif}[2]{\frac{\fdif\!#1}{\fdif\!#2}}
\newcommand{\Laplace}{\boldsymbol{\triangle}}%

\begin{document}
\title{Liquid crystalline growth within a phase-field crystal model}

\author{Sai Tang}
\affiliation{State Key Laboratory of Solidification Processing, Northwestern Polytechnical University, Youyi Western Road 127, 710072 Xi'an, China}
\affiliation{Institut f\"{u}r Wissenschaftliches Rechnen, Technische Universit\"{a}t Dresden, 01062 Dresden, Germany}

\author{Simon Praetorius}
\author{Rainer Backofen}
\affiliation{Institut f\"{u}r Wissenschaftliches Rechnen, Technische Universit\"{a}t Dresden, 01062 Dresden, Germany}

\author{Axel Voigt}
\affiliation{Institut f\"{u}r Wissenschaftliches Rechnen, Technische Universit\"{a}t Dresden, 01062 Dresden, Germany}
\affiliation{Center of Advanced Modeling and Simulations, Technische Universit\"{a}t Dresden, 01062 Dresden, Germany}

\author{Yan-Mei Yu}
\affiliation{Institute of Physics, Chinese Academy of Science, P. O. Box 603, 100190 Beijing, China}

\author{Jincheng Wang}
\affiliation{State Key Laboratory of Solidification Processing, Northwestern Polytechnical University, Youyi Western Road 127, 710072 Xi'an, China}

\begin{abstract}
By using a phase-field crystal (PFC) model, the liquid-crystal growth of the plastic triangular phase is simulated with emphasis on crystal shape and topological defect formation. The equilibrium shape of a plastic triangular crystal (PTC) grown from a isotropic phase is compared with that grown from a columnar/smectic A (CSA) phase. While the shape of a PTC nucleus in the isotropic phase is almost identical to that of a classical PFC model, the shape of a PTC nucleus in CSA is affected by the orientation of stripes in the CSA phase, and irregular hexagonal, elliptical, octagonal, and rectangular shapes are obtained. Concerning the dynamics of the growth process we analyse the topological structure of the nematic-order, which starts from nucleation of $+\frac{1}{2}$ and $-\frac{1}{2}$ disclination pairs at the PTC growth front and evolves into hexagonal cells consisting of $+1$ vortices surrounded by six satellite $-\frac{1}{2}$ disclinations. It is found that the orientational and the positional order do not evolve simultaneously, the orientational order evolves behind the positional order, leading to a large transition zone, which can span over several lattice spacings. 
 
\end{abstract}
\maketitle

\section{Introduction}
Phase field crystal (PFC) models, introduced by Elder et al. \cite{Elder_PRL,Elder_PRE} are today widely used in modeling of crystallization, e.g. \cite{Emmerich,granasy_SM,Tang_AM,Tang_JCG,YU}. As a mean-field approach, the PFC model is capable to describe, over diffusive time scales, atomic arrangements, crystalline defects and interface structures of the crystal growth process. The modeling approach has also been generalized to other systems, e.g. liquid crystals. Liquid crystal PFC (LC-PFC) models \cite{lowen_JPCM} are formulated in terms of three order parameter fields including the reduced translational density $\psi(\vec{\emph{r}},t)$, the local nematic order parameter $S(\vec{\emph{r}},t)$, and the mean nematic director $\hat{n}(\vec{\emph{r}},t)$, wherein $\vec{r}=(x,y)$ and $t$ are position and time, respectively. The couplings among these fields in two dimensions produce the liquid crystalline phases, such as isotropic, nematic, columnar/smectic-A, and plastic crystalline phases (a meso-phase which presents ordered translational density field with crystalline symmetry and macroscopically disordered orientation field \cite{Mock_langmuir, demirors_langmuir, hosein_jmc, rey_jpcb, tkachenko_pre}). Achim et al. \cite{Achim_PRE_2011} numerically determined the stable liquid-crystalline phases and Praetorius et al. \cite{Simon_PRE_2013} found the corresponding phase coexistence regions for special coupling-parameter combinations. The latter group also explored the structure and width of the equilibrium PTC-isotropic and PTC-CSA interface as a function of model parameters related to the coupling strength of the three fields and crystal anisotropy.

We here extend these studies to investigate the growth of a PTC nucleus from the isotropic and the CSA phase. Experimental results \cite{glicksman_jcg, oswald_book} have shown such growth processes to produce a large variety of shapes, e.g. with smooth or faceted dendrites, faceted equilibrium shapes or circular shapes. The growth of Succinonitrile and Pivalic acid \cite{glicksman_jcg} from melt e.g. show high similarity with metal systems in terms of the observed morphologies. This suggests that for some plastic materials the coupling strength of the translational and orientational field may have only limited influence on the morphology. Is this specific to these materials or can the coupling strength also play an important role in the shape evolution of plastic crystal growth? Besides the morphology we will also consider topological defects of the nematic director $\hat{n}(\vec{\emph{r}})$. They have already been  considered e.g. in \cite{Mermin_rmp, Alexander_rmp, Achim_PRE_2011, lowen_epl}. Achim et al. \cite{Achim_PRE_2011} obtained vortices, disclinations, sources or sinks, and hyperbolic points in PTC. Cremer et al. \cite{lowen_epl} depicted the topological defect structure in PTC, and proposed a simplified but topologically equivalent model to explain qualitatively how the topological defect structure of hexagonal symmetry arises. Nevertheless, the understanding about the process of topological defect formation is still limited. How does the topological defect originate from the mother phase? How is a topological defect structure of geometric symmetry constructed?  We will answer these questions by visualizing the topological defect formation on particle scale using the LC-PFC model.

\section{liquid crystal PFC model}

\subsection{Free-energy functional}
We consider a formulation of the LC-PFC model using the reduced translational density field $\psi(\vec{r},t)$, and the symmetric and traceless nematic order tensor fields $Q_{ij}(\vec{r},t)$. In two dimensions the $Q_{ij}(\vec{r},t)$ fields are related to the nematic order parameter $S(\vec{r},t)$ and the nematic director $\hat{n}(\vec{r},t)=(n_{1},n_{2})$ fields through
\begin{equation}
Q_{ij}(\vec{r},t)=S(\vec{r},t)\Big(n_{i}(\vec{r},t)n_{j}(\vec{r},t)-\frac{1}{2}\delta_{ij}\Big).
\label{eq:Q_tensor}
\end{equation}
Thus, the nematic order parameter $S(\vec{r},t)$ and nematic director $\hat{n}(\vec{r},t)$ can be obtained from the nematic order tensor $Q_{ij}(\vec{r},t)$. The dimensionless free energy-functional is written as \cite{lowen_JPCM,Achim_PRE_2011,Simon_PRE_2013},
\begin{equation}%
\!\!\!\!\!\!\!\begin{split}
\mathcal{F}[\psi,Q_{ij}]&=\!\int\!\!\dif^{2}r\,\bigg(\!-\frac{\psi^{3}}{3}+\frac{\psi^{4}}{6}
+(\psi-1)\frac{\psi Q^{2}_{kl}}{4}\\
&\hspace{-5mm}+\frac{Q^{2}_{kl}Q^{2}_{mn}}{64}+A_{1}\psi^{2}
+A_{2}\psi(\Laplace+\Laplace^{2})\psi \\[0.5mm]
&\hspace{-5mm}+B_{3}(\partial_{k}\psi)(\partial_{l}Q_{kl})+D_{1}Q^{2}_{kl}+D_{2}(\partial_{l}Q_{kl})^{2}\bigg),
\end{split}
\label{eq:PFCs}
\end{equation}%
where, Einsteins's sum convention is used, $\Laplace\equiv\partial^{2}_{k}$ is the Laplace operator, and $A_{1}$, $A_{2}$, $B_{3}$, $D_{1}$, and $D_{2}$ are dimensionless model parameters, with $A_{1}$ controling the crystalline anisotropy and $B_{3}$ the coupling strength between $\psi(\vec{r},t)$ and $Q_{ij}(\vec{r},t)$.

\subsection{Dynamic equations}
The dynamic equations of $\psi(\vec{r},t)$ and $Q_{ij}(\vec{r},t)$ are deduced
from classical dynamical density functional theory (DDFT), as written by
\cite{RW_prea, RW_preb},
{\allowdisplaybreaks
\begin{align}%
\begin{split}%
\dot{\psi} + \partial_{i} J^{\psi}_{i} &= 0,
\end{split}\label{eq:PFCdI}\\
\begin{split}
\dot{Q}_{ij} + \Phi^{Q}_{ij}           &= 0,
\end{split}\label{eq:PFCdII}%
\end{align}}%
where $J^{\psi}_{i}(\vec{r},t)$ is the dimensionless current and $\Phi^{Q}_{ij}(\vec{r},t)$ is the dimensionless quasi-current. In constant-mobility approximation, this current and quasi-current are given by \cite{Emmerich}
{\allowdisplaybreaks
\begin{align}%
\begin{split}%
\!\!\!\!\!\!J^{\psi}_{i}  = &- 2\alpha_{1}(\partial_{i}\psi^{\natural})
-2\alpha_{3}(\partial_{j}Q^{\natural}_{ij}),
\end{split}\label{eq:Jpsi}\\[3pt]
\begin{split}%
\!\!\!\!\!\!\Phi^{Q}_{ij} = &-4\alpha_{1}(\Laplace Q^{\natural}_{ij})
-2\alpha_{3}\big(2(\partial_{i}\partial_{j}\psi^{\natural})
-\delta_{ij}(\Laplace\psi^{\natural})\big) \!\!\!\!\!\! \\
&+8\alpha_{4}Q^{\natural}_{ij},
\end{split}\label{eq:PhiQ}%
\end{align}}%
where $\alpha_{1}$, $\alpha_{3}$, and $\alpha_{4}$ are dimensionless mobilities. The thermodynamic conjugates $\psi^{\natural}$ and $Q^{\natural}_{ij}$ are given by
\begin{equation}
\psi^{\natural}=\Fdif{\mathcal{F}}{\psi}\;,\qquad
Q_{ij}^{\natural}=\Fdif{\mathcal{F}}{Q_{ij}},
\end{equation}
which read
{\allowdisplaybreaks%
\begin{align}%
\begin{split}%
\psi^{\natural} = &- \psi^{2} + \frac{2}{3}\psi^{3} + (2\psi-1)\frac{Q^{2}_{ij}}{4} +2A_{1}\psi \\
&+ 2A_{2}(\Laplace+\Laplace^{2})\psi -B_{3}(\partial_{i}\partial_{j}Q_{ij}),
\end{split}\label{eq:psiR}\\%
\begin{split}%
Q^{\natural}_{ij} = &\,\psi(\psi-1)Q_{ij} + \frac{Q_{ij}Q^{2}_{kl}}{8} \\
&-B_{3}\big(2(\partial_{i}\partial_{j}\psi)-\delta_{ij}\Laplace\psi\big) + 4D_{1}Q_{ij} \\[2pt]
&-2D_{2}\:\!\partial_{k}\big(\partial_{i}Q_{kj}+\partial_{j}Q_{ki}-\delta_{ij}(\partial_{l}Q_{kl})\big).
\end{split}\label{eq:QR}%
\end{align}}%
The dynamic equations (3) and (4) will be solved numerically by a semi-implicit Fourier method, see appendix A for details.

\section{Results and discussion}

We consider first a PTC nucleus in the isotropic and CSA phase and show that the coupling
strength between the nematic order tensor and the density, $B_3$, has only a minor influence
on the crystal morphology. Secondly, the ordering of the nematic order tensor during growth of a
PTC into the CSA phase is studied. Independent on the coupling strength, $B_3$, +1 disclinations 
in the PTC phase are formed by coalescence of two
$+\frac{1}{2}$ disclinations. Only the time evolution and the delay of the defect formation with respect to the interface velocity is dependent on $B_3$.

\subsection{Stationary Interfaces}

All simulations are carried out with $A_2=14$, $D_1=1$ and $D_2=0.8$. 
\begin{figure}[htbp]
\begin{center}
\includegraphics[width=5cm]{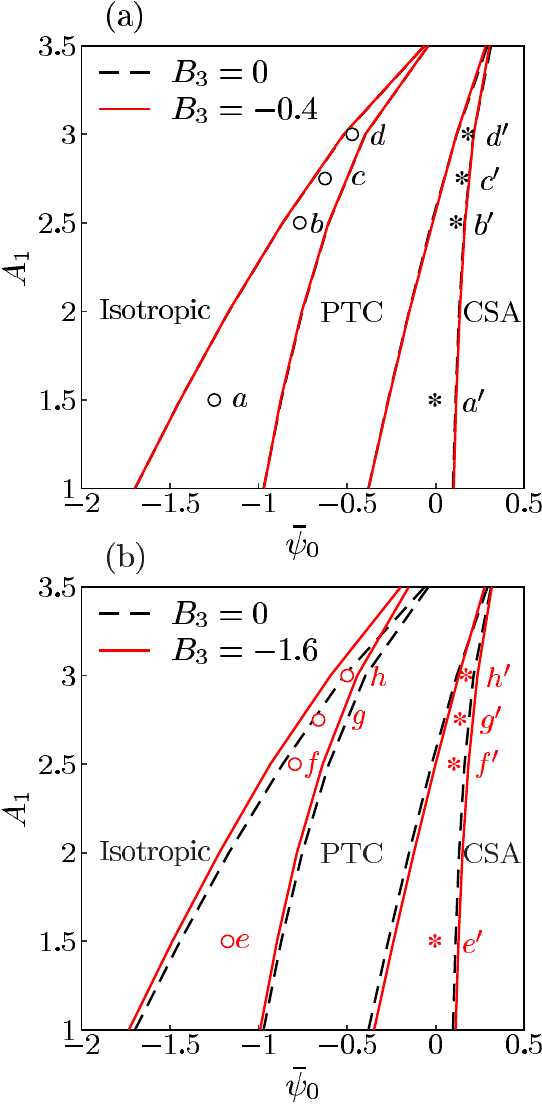}
\end{center}
\vspace{-5mm}
\caption{LC-PFC phase diagrams as function of $\bar{\psi}_{0}$ and
  $A_{1}$ for $B_{3}=0, -0.4$ (a), and $-1.6$ (b) that correspond to zero, weak, and
  strong coupling strength between the density field and the nematic-order
  field, respectively. The small letters correspond to parameters used in our 
  simulations below. \label{fig:PD}}
\end{figure}
\begin{figure*}[htbp!]
\begin{center}
\includegraphics[width=15.7cm]{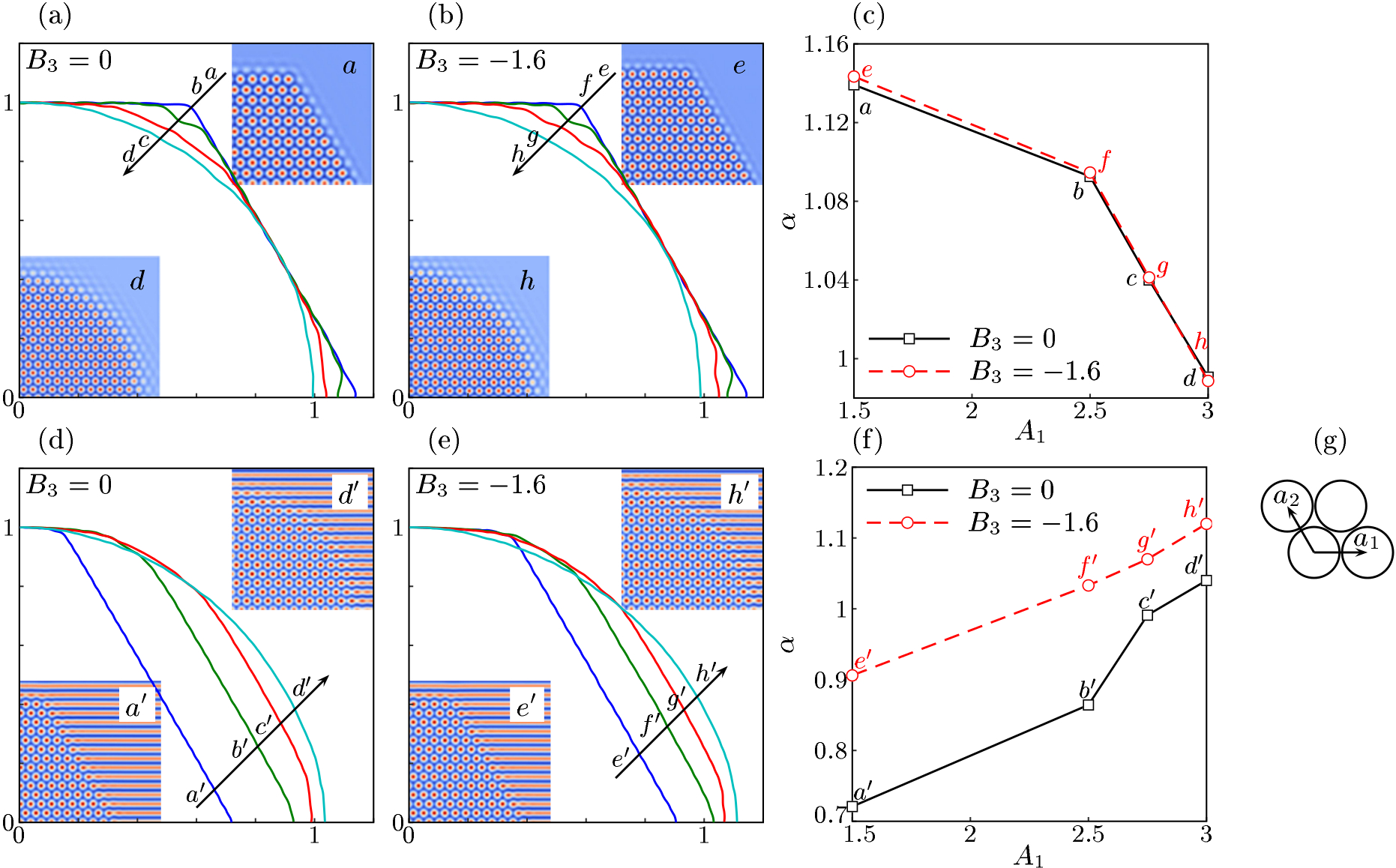}
\end{center}
\vspace{-5mm}
\caption{The equilibrium shape of the PTC nucleus in the isotropic phase (top) for
  $B_{3}=0$ (a) and $B_{3}=-1.6$ (b) and the corresponding contour
  line and aspect ratio $\alpha$ as a function of $A_1$ in (c). The considered examples a-d and e-h 
  correspond to the depicted points in Fig. \ref{fig:PD}. The inlets show the obtained equilibrium 
  shapes for two selected points. The morphology of a PTC nucleus in a CSA phase 
  (bottom) for $A_{1}=1.5$, where the close-packed direction of the PTC phase is parallel 
  to the stripe in the CSA phase. (d) show the equilibrium PTC shapes for $B_{3}=0$ for the considered examples a'-d',   
  (e) shows the shapes of growing PTC for $B_3=-1.6$ for h'-e', again corresponding to the depicted points in Fig.  
  \ref{fig:PD}. The inlets again show selected growth shapes. (f) shows the aspect ratio  $\alpha$ as a function of $A_1$ 
  for the equilibrium PTC shape for $B_{3} = 0$ and the growing shape for $B_{3}=-1.6$. (g) shows the definition of
  crystal unit vectors $a_1$ and $a_2$.} \label{fig:PTC-ISO-CSA}
\end{figure*}
Variations of these parameters turn out to show only minor effects on the considered situation of isotropic, PTC and CSA phases. We vary the mean density $\bar{\psi}_{0}$, anisotropy $A_1$ and coupling
strength $B_3$, which can be used to control the liquid crystal 
phases ~\cite{Achim_PRE_2011, Simon_PRE_2013}.

In Fig.~\ref{fig:PD}, the phase diagram without coupling to the nematic tensor,
$B_3=0$, is compared to the phase diagrams with weak and strong coupling,
$B_3=-0.4$~resp.~$-1.6$.     
The structure of the phase diagrams reflect that of the classical PFC, 
where the PTC and the CSA phase correspond to the crystal and
stripe phase, cf.~\cite{Elder_PRL}.  Thus, analog to the classical PFC $A_1$ is
connected to the undercooling. 
The coupling strength does not change the phase diagram qualitatively. Small
coupling, $B_3=-0.4$, shows no influence, only strong
coupling, $B_3=-1.6$, slightly increases the region of a stable PTC phase.     
Thus, the coexistence region of a PTC-isotropic and a PTC-CSA phase is nearly independent
on $B_3$ and allows us to study the interface properties dependent on coupling
strength,~$B_3$, only. In order to minimize boundary effects in our simulations, a single PTC nucleus is
considered in the center of the simulation domain. To neglect induced stresses by the boundaries, the 
parameters $A_1$ and $B_3$ are
varied and the corresponding $\bar{\psi}_{0}$ value is chosen carefully to assure
coexistence  with the surrounding phase, cf.~\cite{Simon_PRE_2013}.
As long as we are in the coexistence
region of the phase diagram, variations in $\bar{\psi}_{0}$ do not change the
morphology but
determines the size of the PTC nucleus. Size dependence for very small nuclei, as 
observed in \cite{Backofen} can here be neglected as only large enough PTC crystals are considered. 
The domain size is at least 2 times bigger than the shown PTC crystal for the
coexistence with the CSA phase and possible size
dependencies are regularly checked by enlarging the simulation box and comparing the results.  

\subsubsection{Equilibrium shape of PTC in isotropic phase}

The equilibrium shape of a PTC nucleus in the isotropic phase is simulated for
increasing $A_1$. Simulation results and parameters are summarized in Fig.~\ref{fig:PTC-ISO-CSA} (a)-(c).
For every equilibrium shape the interface has been extracted and normalized by
the extension of the PTC nucleus in [12]-direction. Without coupling, $B_3=0$,
the shape changes from a perfect hexagon, $A_1=1.5$, to a circle, $A_1=3$. The
width of the interface seen in the inlet of Fig.~\ref{fig:PTC-ISO-CSA} (a) widens
for increasing $A_1$. This corresponds to the findings in classical PFC for
decreasing undercooling, cf. \cite{Backofen}. Coupling to the nematic tensor
does not change the picture at all, see Fig.~\ref{fig:PTC-ISO-CSA} (b). The
anisotropy of the nucleus is quantified by the reduced aspect ratio
$\alpha$ (ratio of the nucleus extension in [10]- and [12]-direction). 
Based on the Wulff construction, the aspect ratio
$\alpha$ relates to the anisotropy of the line energy~\cite{Khare_SS, Backofen}.
Thus, the nematic order has minor influence on the anisotropy of the line energy
of the PTC phase in the isotropic phase and the anisotropy is well controlled by
$A_1$ only and reflects the results of classical PFC~\cite{Backofen}.

\subsubsection{Crystal shape of PTC in CSA phase}

For PTC growth in the CSA phase, the orientation of the nucleus with respect to the stripes
of the CSA phase becomes important. We consider two set ups. Firstly, the
[12]-direction of the PTC crystal is perpendicular to the stripes. That is,
closed packed layers of the crystal and stripes of the CSA phase are aligned
perfectly (case I).  Secondly, the crystal is rotated by $\frac{\pi}{2}$ and
the stripes do not fit the particle layers (case II).    

The simulation results for case I are shown in  Fig.~\ref{fig:PTC-ISO-CSA} (d)-(f). 
As before, the growth of a small PTC nucleus is simulated until a steady state
is reached. The anisotropy of the steady state is again controlled by
$A_1$. Increasing $A_1$ changes the morphology from faceted to round,~Fig.~\ref{fig:PTC-ISO-CSA} (d). 
But the round shapes are here elongated along the stripes, while the
faceted shapes are elongated perpendicular to the stripes.      
Unlike growth into the isotropic phase, the
growth kinetics is very anisotropic for small $A_1$. That is, the growth of the 
crystal facet in [12]-direction is faster than the growth of the
[21]-facet. 
This leads to a smaller [12]-facet compared to the [21]-facet 
of the stationary crystal. The ratio of the length of two facets is
dependent on initial condition and domain size of the simulation. Thus, they
are just metastable or frozen states and do not represent the overall anisotropy of 
the interface energy. But, as there has been always facets, we can conclude
that the line energy has strong minima for [12]- and [21]-facets, but we
cannot judge the energy ratio between them.
   
Increasing the coupling strength, $B_3$, the transition from faceted to round
shapes does not change, but the elongation of the round shapes along the
stripes is slightly increased, see~Fig.~\ref{fig:PTC-ISO-CSA} (e).  
The coupling strength in this situation is more important than for the
PTC-isotropic interface. 

\begin{figure}[htbp!]
\begin{center}
\includegraphics[width=8cm]{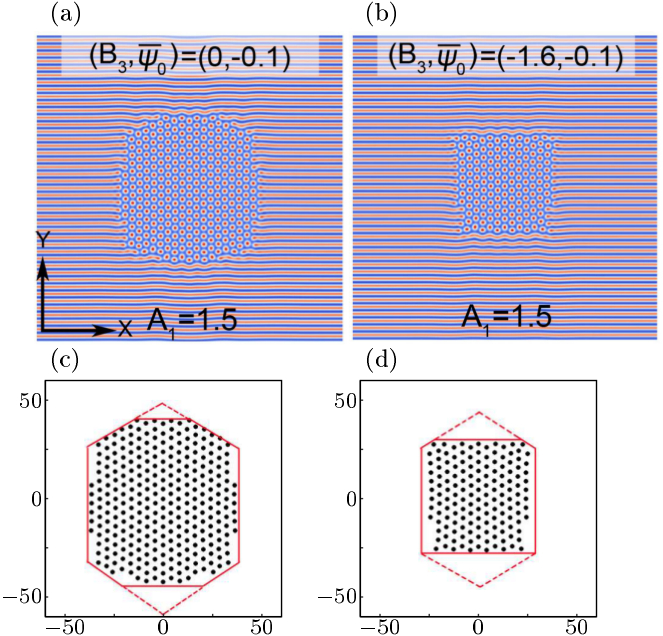}
\end{center}
\vspace{-5mm}
\caption{The PTC shape in the CSA phase for $A_{1}=1.5$, $B_{3}=0$ (a) and (c) and
  $B_{3}=-1.6$ (b) and (d), where the closed packed crystal direction is
  perpendicular to the stripe in the CSA phase. (c) and (d) show a sketch of the PTC
  shape in CSA phase. \label{fig:PTC-CSAII}}
\end{figure}
In case II, the crystal is rotated and the simulations are done as before. 
The PTC crystal does not fit anymore into the layers of the CSA phase, 
see Fig.~\ref{fig:PTC-CSAII}. In this configuration only small crystals for small $A_1$ could be stabilized. Various attempts to
increase nucleus sizes lead to rotating or vanishing crystals. 
The [10]-plane of the PTC crystal is  parallel to the stripes and stabilized
by the CSA phase.  This leads roughly to a hexagon with the edges in
[10]-direction 
cut. The CSA phase is also distorted near the interface. If the
stripes do not fit to the structure of the PTC crystal, the crystal and the CSA
phase is
inhomogeneously strained. This is clearly shown by the bending of the closed packed
particle layers in Fig.~\ref{fig:PTC-CSAII}~(b)~ and the waviness of the
stripes in the CSA phase. The coupling strength
increases the discussed 
features and lead to nearly rectangular shapes,
Fig.~\ref{fig:PTC-CSAII}~(b,d). Thus, case~II is 
energetically penalized by the elastic stress in the crystal induced by 
the surrounding CSA phase and will most likely not occur in liquid crystal
phase transition. In the following we will concentrate on case~I.
  
\subsection{Topological defect formation}

\begin{figure}[htbp!]
\begin{center}
\includegraphics[width=8.5cm]{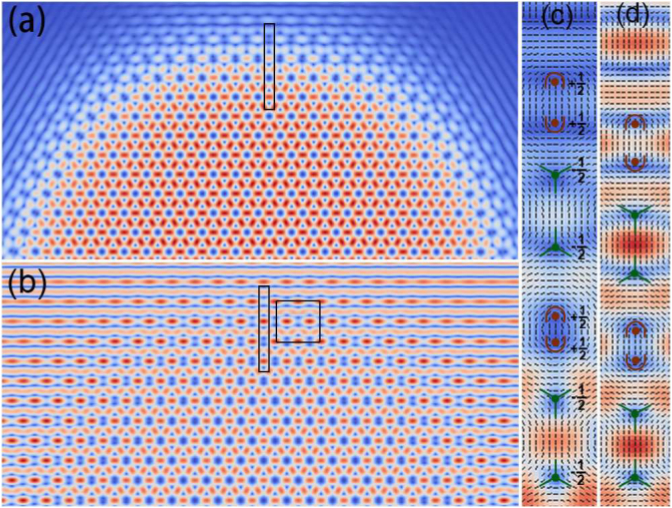}
\end{center}
\vspace{0mm} \caption{The topological structure of the PTC phase grown from
  the isotropic phase (a) and the CSA phase (b). The magnified view of area
  denoted by black boxes in (a) and (b) is given in (c) and (d), respectively,
  where the short arrows represent the director field of the topological
  structure.} \label{fig:top-def-1}
\end{figure}

\begin{figure}[btp!]
\begin{center}
\includegraphics[width=8.5cm]{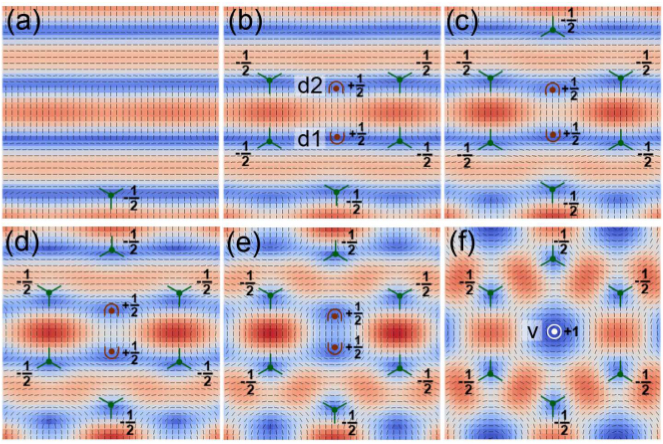}
\end{center}
\vspace{-5mm} \caption{The snapshots of the topological defect structure
  formation in the region enclosed by square box in Fig. \ref{fig:top-def-1} (b) during the PTC
  growth in the CSA phase. d1 and d2 are a pair of $+ \frac{1}{2}$ disclinations. 
  The PTC-CSA interface moves from bottom to top from (a) to (f), and
  for (d) and (e) the region around d1 and d2 is locating inside the PTC-CSA
  interface.} \label{fig:top-def-2}
\end{figure}

The PTC phase exhibits not only the  characteristics of
the crystalline phase, but also shows nematic ordering with topological
defects. Here we examine, how the nematic ordering evolves at the PTC-isotropic
and the PTC-CSA interface.    
The simulations are set up as above. But we restrict ourselves to the
interface in the regime of  constant growth velocity. 
Fig.~\ref{fig:top-def-1} shows the nematic order parameter $S(\vec{r})$ and director
$\hat{n}(\vec{r})$ at the [12]-interface. In the PTC phase along the shown
direction, $+1$ vortices and pairs of $-\frac{1}{2}$ disclinations alternate~\cite{Achim_PRE_2011}. 
Towards the isotropic phase the nematic order parameter not only decreases, but also the
structure of the ordering changes, see Fig.~\ref{fig:top-def-1} (a) and (c) and for details Fig. ~\ref{fig:top-def-2}. 
The $+1$ vortex splits in two $+\frac{1}{2}$ disclinations. This
disclination pair increases its distance and finally vanishes as the 
nematic ordering vanishes. Growth towards the CSA phase, changes the situation 
only a little, Fig.~\ref{fig:top-def-1} (b) and (d). 
The PTC and CSA phase show different nematic
ordering, but the splitting of $+1$ vortices is also observed.
Furthermore, in the PTC isotropic growth the nematic order parameter forms a
weak columnar structure ahead of the growth front. Thus, the PTC isotropic
phase transition has an intermediate stripe phase.
The structure of the topological defects of the nematic director
$\hat{n}(\vec{r},t)$ at the PTC growth front are similar for both cases. 

Fig.~\ref{fig:top-def-2} shows the
formation process of the topological defect structure of the PTC
phase grown towards the CSA phase. Firstly, fluctuation in the $\hat{n}(\vec{r},t)$ field and nematic
order parameter field $S(\vec{r},t)$ arises up at the
PTC growth front. This leads to $+\frac{1}{2}$ and
$-\frac{1}{2}$ disclination pairs, Fig. \ref{fig:top-def-2} (a) and (b). Next, two $+\frac{1}{2}$ disclinations
approach each other gradually and coalescence into a $+1$ vortex, Fig. \ref{fig:top-def-2}
(c) and (e). The $-\frac{1}{2}$ disclinations do not move. At the end, a basic 
unit, the $+1$ vortex surrounded by six $-\frac{1}{2}$ disclinations is
formed, Fig. \ref{fig:top-def-2} (f). 
This constitutes the topological defect structure in the bulk
PTC phase. It was found that the coalescence of two disclinations with the
same charge is energetically unfavored \cite{Lopez-Leon_pre}. Nevertheless, 
during PTC growth, the advancing interface drives coalescence of two $+\frac{1}{2}$
disclinations. In the following, the kinetics of this topological defects
formation will be discussed in more detail. 
\begin{figure}[htbp!]
\begin{center}
\includegraphics[width=8.5cm]{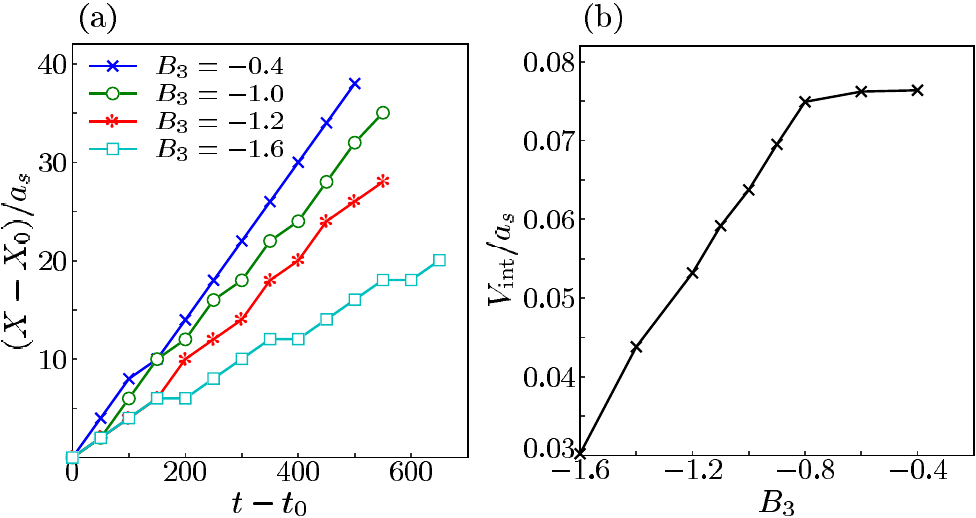}
\end{center}
\vspace{-5mm} \caption{The growth kinetics of PTC-CSA interface. The displacement of PTC-CSA interface versus time $t$ (a), and
  growth velocity (b) for various $B_{3}$. $X$ is the position of PTC-CSA
  interface at different times, and $X_{0}$ the initial position at time
  $t_{0}$. \label{fig:interface-vel}}
\end{figure}
Firstly, we study the growth velocity of the PTC nucleus from the CSA
phase along the direction normal to stripes as shown in Fig. \ref{fig:top-def-1} (b). After some
initial relaxation the displacement of the PTC-CSA interface increases linearly
with time and a constant growth velocity, $V_{int}$, is achieved,
Fig.~\ref{fig:interface-vel} (a). The dependence of the growth velocity on the
coupling strength $B_3$ is
shown in Fig.~\ref{fig:interface-vel} (b). A strong and weak coupling regime can
be identified. 
For strong coupling, 
the growth velocity increases linearly with $B_{3}$, changing from -1.6 to
-0.8. For weak coupling, when $B_{3}$ is in the range between $-0.8$ and $-0.4$,
the growth velocity shows little dependence on $B_{3}$. Starting from the
growth velocity $V_{int}$, the defect moving velocity $V_{d}$ is scaled as
$\widetilde{V}=V_{d}/V_{int}$, and time $t$ is scaled as
$\widetilde{t}=t/(a_s/V_{int})$, with the lattice constant of the CSA phase $a_s=2\pi/\sqrt{1/2} \approx
  8.8858$. This can be used to count the number of
lattices in the CSA phase that the PTC-CSA interface has advanced.
\begin{figure}[htbp!]
\begin{center}
\includegraphics[width=8.5cm]{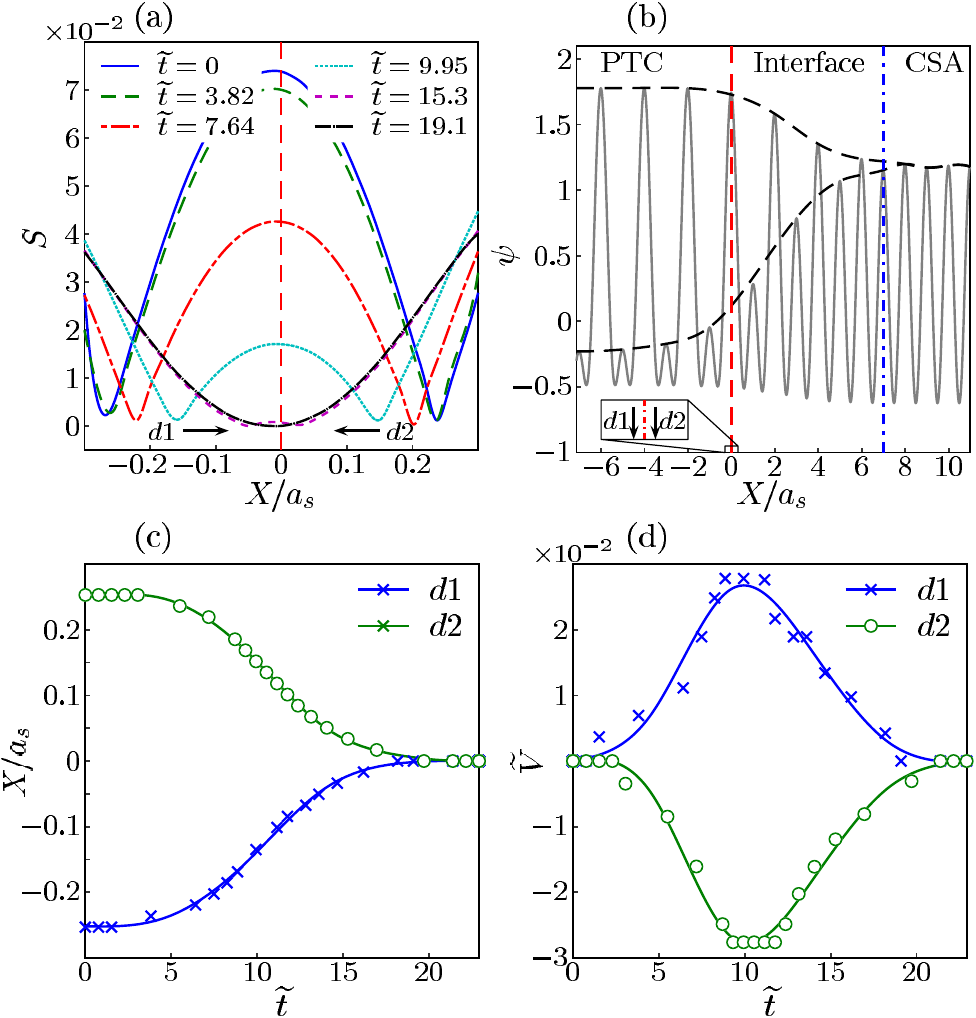}
\end{center}
\vspace{-5mm} \caption{The moving trajectory of two $+ \frac{1}{2}$ disclinations d1 and d2, corresponding to Fig. \ref{fig:top-def-2}, as represented by the $S(\vec{r},t)$ profile (a), profile of density field corresponding to the $S(\vec{r},t)$ when $t=130$ (i.e., $\widetilde{t}=12.5$) (b), the position (c), and the velocity (d) of d1 and d2. Note that the red dashed line in (a) and (b) denotes the region in the vicinity of d1 and d2, and the region on the left side of the red dashed line in (b) is the PTC phase, the region on the right sides of the blue dash doted line is the CSA phase, and the region between the two lines is the PTC-CSA interface (the left and the right boundary of the interface is determined by the criterion that the height of the peaks is lower than 0.95 of that in the bulk PTC phase and the CSA phase, respectively). The arrows in (a) denote the moving direction of d1 and d2.} \label{fig:defect-vel-detail}
\end{figure}
The coalescence of two $+\frac{1}{2}$ disclinations is illustrated in
Fig. \ref{fig:defect-vel-detail} in more detail. The position of the topological defects are easily
identified at the minima of the nematic order parameter $S(\vec{r})$, Fig. \ref{fig:defect-vel-detail} (a).  
Initially, there are two minima in $S(\vec{r},t)$ profiles
which correspond to the disclinations d1 and d2, respectively. The two minima approach
gradually and coalesce finally at $\widetilde{t}={19.1}$. 
To compare the evolution of the density field and
the orientational field, we extract the density profile when the PTC-CSA interface
is sweeping the region around d1 and d2 at $\widetilde{t}=12.5$. 
As shown in Fig. \ref{fig:defect-vel-detail} (b), the peak value
of the density profile at the position marked by the left red dash line almost
equals that of the bulk PTC phase. However, the corresponding
$S(\vec{r},t)$ profile at the same time ($\widetilde{t}=12.5$) indicates that 
the evolution of the orientational field in the region
around d1 and d2 is still far from being completed. The distance between d1
and d2 versus $\widetilde{t}$ is illustrated in Fig. \ref{fig:defect-vel-detail} (c). We can see that d1
and d2 move nearly symmetrically. The velocities of d1 and d2 are given in
Fig. \ref{fig:defect-vel-detail} (d). The largest velocity occurs about the time $\widetilde{t}=12.5$.
This corresponds to the moment when the growth front is sweeping the
region around d1 and d2, which corresponds to the state in Fig. \ref{fig:top-def-2} (e), and when the density field
evolution around d1 and d2 has almost completed, as shown in Fig. \ref{fig:defect-vel-detail} (b). With the
PTC-CSA interface approaching the region around d1 and d2, the velocities
of d1 and d2 increase rapidly,
while decrease steeply after the interface left away from them. In other
words, the evolution of topological defect formation is accelerated by the
advancing PTC-CSA interface. When the $+1$ vortex forms through the
coalescence of d1 and d2, the interface has moved forward by a distance of
about 20 lattice constants compared with the position when it passes d1 and
d2. 
\begin{figure}[htbp!]
\begin{center}
\includegraphics[width=8.5cm]{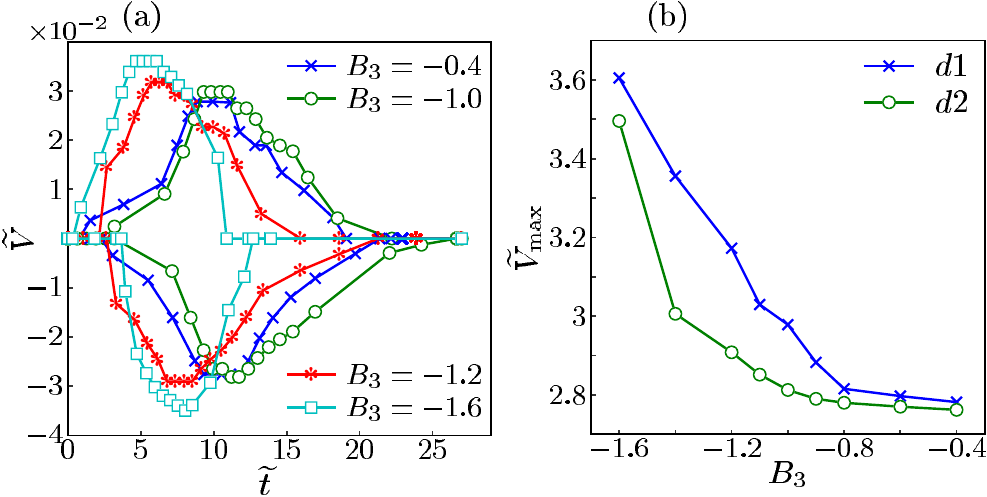}
\end{center}
\vspace{-5mm} \caption{The influences of coupling strength $B_3$ on
  topological defects formation kinetics: (a) the velocity of two $+
  \frac{1}{2}$  disclinations during the formation of $+1$ vortex for various
  $B_{3}$, and (b) the peak value of the velocity curves as shown in (a) as a
  function of $B_{3}$.} \label{fig:defect-vel-B}
\end{figure}

Finally, we investigated the influence of coupling strength on $+1$ vortex
formation kinetics. As shown in Fig. \ref{fig:defect-vel-B} (a), with coupling strength increasing
by decreasing $B_{3}$ from -0.4 to -1.6, the peaks of velocity curves
increase, and the scaled time $\widetilde{t}$ needed for the formation of a
$+1$ vortex decreases. This indicates that the evolution from d1 and d2 to a
$+1$ vortex can be completed faster for higher coupling strength. The dependence 
of peak velocities of
the curves in Fig. \ref{fig:defect-vel-B} (a) on $B_{3}$ is summarized in Fig. \ref{fig:defect-vel-B} (b). For weak
coupling strength when $B_{3}$ is between $-0.8$ and $-0.4$, the peak velocities of d1
and d2 are nearly the same, and the motion of d1 and d2 are approximately
symmetrical. However, with coupling strength increasing further ($B_{3}<-1.0$),
the peak velocities of d1 and d2 increase substantially, and d1,
the disclination moving along the same direction as the PTC-CSA interface advances, 
has a larger peak velocity than that of d2, moving opposite to the
PTC-CSA interface. Thus, it is shown that the increase of coupling strength
accelerates  the topological defect formation during PTC growth. 

\section{Conclusion}
In summary, by using the LC-PFC model we have
investigated the growth of PTC nucleus from the isotropic phase and the CSA phase on
particle scale. An overall picture for the growth of the PTC phase is presented for shape evolution 
and nematic topological defect structure formation. It is demonstrated that the shape evolution for the
PTC phase growth are mainly
determined by crystalline anisotropy. The coupling
strength exerts little influence on the shape of growing PTC
nucleus. Only the shape of the PTC nucleus grown from the CSA phase also
depends on the misorientation of PTC and CSA. Moreover, for the formation
process of the nematic topological structure of the PTC phase, the formation
of the PTC topological structure starts from nucleation of $+\frac{1}{2}$ and
$-\frac{1}{2}$ disclination pairs at the PTC growth front, and leads to
the coalescence of $+\frac{1}{2}$ pairs forming a
hexagonal cell consisting of one $+1$ vortex surrounded by six $-\frac{1}{2}$
disclinations. The coupling strength influences the kinetics of
topological defect formation, and strong coupling strength accelerates the
formation of nematic topological structure. Thus, while morphological shapes of plastic crystals
might look similar to that of metal systems, the dynamics of the growth process shows strong differences.

For future work, it is interesting and meaningful to extend the present study to three spatial dimensions. The topological defect formation of plastic crystal with crystal structures, such as simple cubic, body-centered crystal, face-centered crystal, etc., may provide more amazing scenarios of topological defect formation. Also, it is worth to study the plastic crystal growth with polar and non-spherical particles.

\begin{acknowledgments}
This work has been supported by EU FP7, PHASEFIELD under no.247504, German Science Foundation within SPP1296 (Grant No. Vo899/7), Nature Science Foundation of China (Grant Nos.10974228, 51071128), and 2013 DAAD-CSC postdoctoral scholarship.
\end{acknowledgments}

\appendix
\section{Semi-implicit Fourier method for the LC-PFC model}
A system of six coupled nonlinear partial differential equations need to be solved in the liquid crystal PFC model. In order to numerically solve this system efficiently, we decoupled and linearized it. Considering the tracelessness and symmetry of the nematic tensor as shown in equation (1), we can extract the variables $q_{i}\equiv Q_{i,1}$ and $q_{i}^{\natural}\equiv Q_{i,1}^{\natural}$, with $\mathbf{q}=(q_1,q_2)^\top$ and $\textbf{q}^\natural=(q_1^\natural,q_2^\natural)^\top$. The dynamic equations \eqref{eq:PFCdI} and \eqref{eq:PFCdII} thus have the compact form
\begin{eqnarray}\label{eq:pde_liquid_crystal_pfc}
\dot{\psi}&=&2\alpha_{1}\triangle \psi^{\natural}+2\alpha_{3}\blacktriangle_{i} q_{i}^{\natural} \nonumber \\
\dot{q}_{i}&=&4\alpha_{1}\triangle q_{i}^{\natural}-8\alpha_{4}q_{i}^{\natural}+2\alpha_{3}\blacktriangle_{i} \psi^{\natural}.
\end{eqnarray}
The terms containing the operator $\blacktriangle$ can be expanded, as
\begin{equation*}
\blacktriangle_{i} q_{i}^{\natural}=(\partial_{1}\partial_{1}-\partial_{2}\partial_{2})q_{1}^{\natural}+2\partial_{1}\partial_{2}q_{2}^{\natural}
\end{equation*}
and the thermodynamic conjugates read
\begin{eqnarray}\label{eq:thermodynamic_conjugates}
\psi^{\natural}&=&\omega_{\psi}(\psi,\textbf{q})+\overbrace{2A_{1}\psi+2A_{2}(\triangle+\triangle^{2})\psi}^{L_{\psi}^\psi(\psi)}+\overbrace{(-B_{3})\blacktriangle_{i}q_{i}}^{L_{\psi}^q(\textbf{q})},\nonumber \\
q_{i}^{\natural}&=&\omega_{q}(\psi,\textbf{q})_{i}+\underbrace{4D_{1}q_{i}-2D_{2}\triangle q_{i}}_{L_{q}^q(q_i)}+\underbrace{(-B_{3})\blacktriangle_{i}\psi}_{L_{q,i}^\psi(\psi)},
\end{eqnarray}
with the linear parts $L_{\psi}^\psi(\psi), L_{\psi}^q(\textbf{q}), L_{q}^q(q_i)$ and $L_{q,i}^\psi(\psi)$ and the polynomials
\begin{eqnarray}\label{eq:nonlinear_polynomials}
\omega_{\psi}(\psi,\textbf{q})&=&-\psi^{2}+\frac {2}{3}\psi^{3}+\frac{1}{2}(2\psi-1)(q_{1}^{2}+q_{2}^{2}), \nonumber \\
\omega_{q}(\psi,\textbf{q})_{i}&=&\psi(\psi-1)q_{i}+\frac {1}{4}q_{i}(q_{1}^{2}+q_{2}^{2}).
\end{eqnarray}
We define the Fourier wave vector $\textbf{k}=(k_1,k_2)^\top$ and introduce the Fourier transform $\mathcal{F}$ of the order parameter fields, as
\[\psi\mapsto\mathcal{F}(\psi)=:\hat{\psi}(\textbf{k}),\quad q_i\mapsto\mathcal{F}(q_i)=:\hat{q_i}(\textbf{k}).\]
Thus, we can write the differential operators in Fourier space, as
\begin{eqnarray}
\triangle\rightarrow \hat{\triangle}&:=&-(k_{1}^{2}+k_{2}^{2}) = -|\textbf{k}|^2 \nonumber \\
\blacktriangle\rightarrow \hat{\blacktriangle}&:=&\begin{pmatrix}
-(k_{1}^{2}-k_{2}^{2}) \\
-k_{1}k_{2}
\end{pmatrix}.
\end{eqnarray}

Discretizing \eqref{eq:pde_liquid_crystal_pfc} in time using a semi-implicit backward Euler-discretization and transforming the equation to Fourier space leads to the spectral method used in our calculations. Therefore, we transform the linear differential operators $L_\ast^\ast$ of \eqref{eq:thermodynamic_conjugates} using the operators defined above and denote it with a hat, i.e.
\begin{eqnarray}\label{eq:transformed_linear_operator}
\hat{L}_{\psi}^\psi[\hat{\psi}] &:=& 2A_1\hat{\psi} + 2 A_2(\hat{\triangle} + \hat{\triangle}^2)\cdot\hat{\psi} \nonumber \\
&=&\big(2A_1 + 2 A_2(-|\textbf{k}|^2 + |\textbf{k}|^4)\big)\cdot\hat{\psi}
\end{eqnarray}
and the other operators in an analogous manner.

The nonlinear polynomials \eqref{eq:nonlinear_polynomials} are evaluated in real space and transformed afterwards, i.e.
\begin{equation}\label{eq:tranformed_nonlinear_operator}
\hat{\omega}_\psi(\psi,\textbf{q}) := \mathcal{F}(\omega_\psi(\psi,\textbf{q})).
\end{equation}
Therefore, in each time step the order parameters $\psi$ and $\mathbf{q}$ have to be transformed between real and Fourier space using efficient implementations of $\mathcal{F}$ and $\mathcal{F}^{-1}$.

Let $0=t^0<t^1<\ldots<t^N=T$ be a discretization of a time interval $[0,T]$, with $t^{n+1}-t^n=:\tau$ the timestep width. The finite difference approximation
\[\dot{\hat{\psi}} \approx \frac{\hat{\psi}^{n+1} - \hat{\psi}^n}{\tau},\]
with $\hat{\psi}^n\equiv \hat{\psi}(t^n)$ and $\dot{\hat{q}}_i$ in an analogous manner, inserted into the transformed equations, using \eqref{eq:transformed_linear_operator} and \eqref{eq:tranformed_nonlinear_operator}, leads to an iterative procedure:

Let $\hat{\psi}^0=\mathcal{F}(\psi^0),\; \hat{\textbf{q}}^0=\mathcal{F}(\textbf{q}^0)$. For $n=0,1,2,\ldots,N$
\begin{enumerate}
\item solve \begin{eqnarray}\label{eq:fourier_discretization_psi}
\Big(1 - \tau &\big[& 2\alpha_1\hat{\triangle}\cdot \hat{L}_{\psi}^\psi + 2\alpha_3\hat{\blacktriangle}_i\cdot \hat{L}_{q,i}^\psi\big] \Big)\hat{\psi}^{n+1}\nonumber \\
=\Big(\hat{\psi}^{n}+\tau &\big[& 2\alpha_1\hat{\triangle}\cdot\big(\hat{L}_{\psi}^q[\hat{\textbf{q}}^n] + \hat{\omega}_\psi(\psi^n,\textbf{q}^n) \big) \nonumber \\
&+& 2\alpha_3\hat{\blacktriangle}_i\cdot\big(\hat{L}_{q}^q[\hat{q}_i^n] + \hat{\omega}_q(\psi^n,\textbf{q}^n)_i\big)\big]\Big),
\end{eqnarray}
\item transform to real space: \[\psi^{n+1} = \mathcal{F}^{-1}(\hat{\psi}^{n+1}),\]
\item for $i=1,2$ solve \begin{eqnarray}\label{eq:fourier_discretization_q}
\Big(1 - \tau &\big[& (4\alpha_1\hat{\triangle} - 8\alpha_4)\cdot \hat{L}_{q}^q\big] \Big)\hat{q}_i^{n+1}\nonumber \\
-\tau &\cdot & 2\alpha_3\hat{\blacktriangle}_i\cdot \hat{L}_{\psi}^q[\hat{\textbf{q}}_{(i)}^{n,n+1}]\nonumber \\
=\Big(\hat{q}_i^{n}+\tau &\big[& (4\alpha_1\hat{\triangle} - 8\alpha_4)\cdot\big(\hat{L}_{q,i}^\psi[\hat{\psi}^{n+1}] + \hat{\omega}_q(\psi^{n+1},\textbf{q}^n)_i \big) \nonumber \\
&+& 2\alpha_3\hat{\blacktriangle}_i\cdot\big(\hat{L}_{\psi}^\psi[\hat{\psi}^{n+1}] + \hat{\omega}_\psi(\psi^{n+1},\textbf{q}^n)\big)\big]\Big),
\end{eqnarray}
where $\hat{\textbf{q}}_{(i)}^{n,n+1}$ are intermediate vectors defined as
\[\hat{\textbf{q}}_{(1)}^{n,n+1} := \begin{pmatrix}
\hat{q}_1^{n+1} \\ \hat{q}_2^{n}
\end{pmatrix},\quad\hat{\textbf{q}}_{(2)}^{n,n+1} := \begin{pmatrix}
\hat{q}_1^{n} \\ \hat{q}_2^{n+1}
\end{pmatrix},\]
\item transform to real space: \[q_i^{n+1} = \mathcal{F}^{-1}(\hat{q}_i^{n+1}),\; i=1,2.\]
\end{enumerate}

In equation \eqref{eq:fourier_discretization_q} the updated values $\psi^{n+1}$ and $\hat{\psi}^{n+1}$ can be used.

\bibliographystyle{apsrev4-1}
\bibliography{references}

\begin{thebibliography}{25}%
\makeatletter
\providecommand \@ifxundefined [1]{%
 \@ifx{#1\undefined}
}%
\providecommand \@ifnum [1]{%
 \ifnum #1\expandafter \@firstoftwo
 \else \expandafter \@secondoftwo
 \fi
}%
\providecommand \@ifx [1]{%
 \ifx #1\expandafter \@firstoftwo
 \else \expandafter \@secondoftwo
 \fi
}%
\providecommand \natexlab [1]{#1}%
\providecommand \enquote  [1]{``#1''}%
\providecommand \bibnamefont  [1]{#1}%
\providecommand \bibfnamefont [1]{#1}%
\providecommand \citenamefont [1]{#1}%
\providecommand \href@noop [0]{\@secondoftwo}%
\providecommand \href [0]{\begingroup \@sanitize@url \@href}%
\providecommand \@href[1]{\@@startlink{#1}\@@href}%
\providecommand \@@href[1]{\endgroup#1\@@endlink}%
\providecommand \@sanitize@url [0]{\catcode `\\12\catcode `\$12\catcode
  `\&12\catcode `\#12\catcode `\^12\catcode `\_12\catcode `\%12\relax}%
\providecommand \@@startlink[1]{}%
\providecommand \@@endlink[0]{}%
\providecommand \url  [0]{\begingroup\@sanitize@url \@url }%
\providecommand \@url [1]{\endgroup\@href {#1}{\urlprefix }}%
\providecommand \urlprefix  [0]{URL }%
\providecommand \Eprint [0]{\href }%
\providecommand \doibase [0]{http://dx.doi.org/}%
\providecommand \selectlanguage [0]{\@gobble}%
\providecommand \bibinfo  [0]{\@secondoftwo}%
\providecommand \bibfield  [0]{\@secondoftwo}%
\providecommand \translation [1]{[#1]}%
\providecommand \BibitemOpen [0]{}%
\providecommand \bibitemStop [0]{}%
\providecommand \bibitemNoStop [0]{.\EOS\space}%
\providecommand \EOS [0]{\spacefactor3000\relax}%
\providecommand \BibitemShut  [1]{\csname bibitem#1\endcsname}%
\let\auto@bib@innerbib\@empty
\bibitem [{\citenamefont {Elder}\ \emph {et~al.}(2002)\citenamefont {Elder},
  \citenamefont {Katakowski}, \citenamefont {Haataja},\ and\ \citenamefont
  {Grant}}]{Elder_PRL}%
  \BibitemOpen
  \bibfield  {author} {\bibinfo {author} {\bibfnamefont {K.~R.}\ \bibnamefont
  {Elder}}, \bibinfo {author} {\bibfnamefont {M.}~\bibnamefont {Katakowski}},
  \bibinfo {author} {\bibfnamefont {M.}~\bibnamefont {Haataja}}, \ and\
  \bibinfo {author} {\bibfnamefont {M.}~\bibnamefont {Grant}},\ }\href@noop {}
  {\bibfield  {journal} {\bibinfo  {journal} {Phys. Rev. Lett.}\ }\textbf
  {\bibinfo {volume} {88}},\ \bibinfo {pages} {245701} (\bibinfo {year}
  {2002})}\BibitemShut {NoStop}%
\bibitem [{\citenamefont {Elder}\ and\ \citenamefont
  {Grant}(2004)}]{Elder_PRE}%
  \BibitemOpen
  \bibfield  {author} {\bibinfo {author} {\bibfnamefont {K.~R.}\ \bibnamefont
  {Elder}}\ and\ \bibinfo {author} {\bibfnamefont {M.}~\bibnamefont {Grant}},\
  }\href@noop {} {\bibfield  {journal} {\bibinfo  {journal} {Phys. Rev. E}\
  }\textbf {\bibinfo {volume} {70}},\ \bibinfo {pages} {051605} (\bibinfo
  {year} {2004})}\BibitemShut {NoStop}%
\bibitem [{\citenamefont {Emmerich}\ \emph {et~al.}(2012)\citenamefont
  {Emmerich}, \citenamefont {L\"{o}wen}, \citenamefont {Wittkowski},
  \citenamefont {Gruhn}, \citenamefont {T\'{o}th}, \citenamefont {Tegze},\ and\
  \citenamefont {Gr\'{a}n\'{a}sy}}]{Emmerich}%
  \BibitemOpen
  \bibfield  {author} {\bibinfo {author} {\bibfnamefont {H.}~\bibnamefont
  {Emmerich}}, \bibinfo {author} {\bibfnamefont {H.}~\bibnamefont {L\"{o}wen}},
  \bibinfo {author} {\bibfnamefont {R.}~\bibnamefont {Wittkowski}}, \bibinfo
  {author} {\bibfnamefont {T.}~\bibnamefont {Gruhn}}, \bibinfo {author}
  {\bibfnamefont {G.~I.}\ \bibnamefont {T\'{o}th}}, \bibinfo {author}
  {\bibfnamefont {G.}~\bibnamefont {Tegze}}, \ and\ \bibinfo {author}
  {\bibfnamefont {L.}~\bibnamefont {Gr\'{a}n\'{a}sy}},\ }\href@noop {}
  {\bibfield  {journal} {\bibinfo  {journal} {Advances in Physics}\ }\textbf
  {\bibinfo {volume} {61}},\ \bibinfo {pages} {665} (\bibinfo {year}
  {2012})}\BibitemShut {NoStop}%
\bibitem [{\citenamefont {Tegze}\ \emph {et~al.}(2011)\citenamefont {Tegze},
  \citenamefont {Gr\'{a}n\'{a}sy}, \citenamefont {T\'{o}th}, \citenamefont
  {Douglas},\ and\ \citenamefont {Pusztai}}]{granasy_SM}%
  \BibitemOpen
  \bibfield  {author} {\bibinfo {author} {\bibfnamefont {G.}~\bibnamefont
  {Tegze}}, \bibinfo {author} {\bibfnamefont {L.}~\bibnamefont
  {Gr\'{a}n\'{a}sy}}, \bibinfo {author} {\bibfnamefont {G.~I.}\ \bibnamefont
  {T\'{o}th}}, \bibinfo {author} {\bibfnamefont {J.~F.}\ \bibnamefont
  {Douglas}}, \ and\ \bibinfo {author} {\bibfnamefont {T.}~\bibnamefont
  {Pusztai}},\ }\href@noop {} {\bibfield  {journal} {\bibinfo  {journal} {Soft
  Matter}\ }\textbf {\bibinfo {volume} {7}},\ \bibinfo {pages} {1789} (\bibinfo
  {year} {2011})}\BibitemShut {NoStop}%
\bibitem [{\citenamefont {Tang}\ \emph {et~al.}(2012)\citenamefont {Tang},
  \citenamefont {Wang}, \citenamefont {Guo}, \citenamefont {Wang},
  \citenamefont {Yu},\ and\ \citenamefont {Zhou}}]{Tang_AM}%
  \BibitemOpen
  \bibfield  {author} {\bibinfo {author} {\bibfnamefont {S.}~\bibnamefont
  {Tang}}, \bibinfo {author} {\bibfnamefont {Z.~J.}\ \bibnamefont {Wang}},
  \bibinfo {author} {\bibfnamefont {Y.~L.}\ \bibnamefont {Guo}}, \bibinfo
  {author} {\bibfnamefont {J.~C.}\ \bibnamefont {Wang}}, \bibinfo {author}
  {\bibfnamefont {Y.~M.}\ \bibnamefont {Yu}}, \ and\ \bibinfo {author}
  {\bibfnamefont {Y.~H.}\ \bibnamefont {Zhou}},\ }\href@noop {} {\bibfield
  {journal} {\bibinfo  {journal} {Acta Mater.}\ }\textbf {\bibinfo {volume}
  {60}},\ \bibinfo {pages} {5501} (\bibinfo {year} {2012})}\BibitemShut
  {NoStop}%
\bibitem [{\citenamefont {Tang}\ \emph {et~al.}(2011)\citenamefont {Tang},
  \citenamefont {Backofen}, \citenamefont {Wang}, \citenamefont {Zhou},
  \citenamefont {Voigt},\ and\ \citenamefont {Yu}}]{Tang_JCG}%
  \BibitemOpen
  \bibfield  {author} {\bibinfo {author} {\bibfnamefont {S.}~\bibnamefont
  {Tang}}, \bibinfo {author} {\bibfnamefont {R.}~\bibnamefont {Backofen}},
  \bibinfo {author} {\bibfnamefont {J.}~\bibnamefont {Wang}}, \bibinfo {author}
  {\bibfnamefont {Y.}~\bibnamefont {Zhou}}, \bibinfo {author} {\bibfnamefont
  {A.}~\bibnamefont {Voigt}}, \ and\ \bibinfo {author} {\bibfnamefont {Y.~M.}\
  \bibnamefont {Yu}},\ }\href@noop {} {\bibfield  {journal} {\bibinfo
  {journal} {J. Cryst. Growth}\ }\textbf {\bibinfo {volume} {334}},\ \bibinfo
  {pages} {146} (\bibinfo {year} {2011})}\BibitemShut {NoStop}%
\bibitem [{\citenamefont {Yu}\ \emph {et~al.}(2011)\citenamefont {Yu},
  \citenamefont {Backofen},\ and\ \citenamefont {A.Voigt}}]{YU}%
  \BibitemOpen
  \bibfield  {author} {\bibinfo {author} {\bibfnamefont {Y.~M.}\ \bibnamefont
  {Yu}}, \bibinfo {author} {\bibfnamefont {R.}~\bibnamefont {Backofen}}, \ and\
  \bibinfo {author} {\bibnamefont {A.Voigt}},\ }\href@noop {} {\bibfield
  {journal} {\bibinfo  {journal} {J. Cryst. Growth}\ }\textbf {\bibinfo
  {volume} {318}},\ \bibinfo {pages} {18} (\bibinfo {year} {2011})}\BibitemShut
  {NoStop}%
\bibitem [{\citenamefont {L\"{o}wen}(2010)}]{lowen_JPCM}%
  \BibitemOpen
  \bibfield  {author} {\bibinfo {author} {\bibfnamefont {H.}~\bibnamefont
  {L\"{o}wen}},\ }\href@noop {} {\bibfield  {journal} {\bibinfo  {journal} {J.
  Phys.: Condens. Matter}\ }\textbf {\bibinfo {volume} {22}},\ \bibinfo {pages}
  {364105} (\bibinfo {year} {2010})}\BibitemShut {NoStop}%
\bibitem [{\citenamefont {Mock}\ and\ \citenamefont
  {Zukoski}(2007)}]{Mock_langmuir}%
  \BibitemOpen
  \bibfield  {author} {\bibinfo {author} {\bibfnamefont {E.~B.}\ \bibnamefont
  {Mock}}\ and\ \bibinfo {author} {\bibfnamefont {C.~F.}\ \bibnamefont
  {Zukoski}},\ }\href@noop {} {\bibfield  {journal} {\bibinfo  {journal}
  {Langmur}\ }\textbf {\bibinfo {volume} {23}},\ \bibinfo {pages} {8760}
  (\bibinfo {year} {2007})}\BibitemShut {NoStop}%
\bibitem [{\citenamefont {Demir\"{o}rs}\ \emph {et~al.}(2010)\citenamefont
  {Demir\"{o}rs}, \citenamefont {Johnson}, \citenamefont {van Kats},
  \citenamefont {Blaaderen},\ and\ \citenamefont {Imhof}}]{demirors_langmuir}%
  \BibitemOpen
  \bibfield  {author} {\bibinfo {author} {\bibfnamefont {A.~F.}\ \bibnamefont
  {Demir\"{o}rs}}, \bibinfo {author} {\bibfnamefont {P.~M.}\ \bibnamefont
  {Johnson}}, \bibinfo {author} {\bibfnamefont {C.~M.}\ \bibnamefont {van
  Kats}}, \bibinfo {author} {\bibfnamefont {A.~V.}\ \bibnamefont {Blaaderen}},
  \ and\ \bibinfo {author} {\bibfnamefont {A.}~\bibnamefont {Imhof}},\
  }\href@noop {} {\bibfield  {journal} {\bibinfo  {journal} {Langmur}\ }\textbf
  {\bibinfo {volume} {26}},\ \bibinfo {pages} {14466} (\bibinfo {year}
  {2010})}\BibitemShut {NoStop}%
\bibitem [{\citenamefont {Hosein}\ \emph {et~al.}(2009)\citenamefont {Hosein},
  \citenamefont {John}, \citenamefont {Lee}, \citenamefont {Escobedo},\ and\
  \citenamefont {Liddell}}]{hosein_jmc}%
  \BibitemOpen
  \bibfield  {author} {\bibinfo {author} {\bibfnamefont {I.~D.}\ \bibnamefont
  {Hosein}}, \bibinfo {author} {\bibfnamefont {B.~S.}\ \bibnamefont {John}},
  \bibinfo {author} {\bibfnamefont {S.~H.}\ \bibnamefont {Lee}}, \bibinfo
  {author} {\bibfnamefont {F.~A.}\ \bibnamefont {Escobedo}}, \ and\ \bibinfo
  {author} {\bibfnamefont {C.~M.}\ \bibnamefont {Liddell}},\ }\href@noop {}
  {\bibfield  {journal} {\bibinfo  {journal} {J. Mater. Chem.}\ }\textbf
  {\bibinfo {volume} {19}},\ \bibinfo {pages} {344} (\bibinfo {year}
  {2009})}\BibitemShut {NoStop}%
\bibitem [{\citenamefont {Rey}(2008)}]{rey_jpcb}%
  \BibitemOpen
  \bibfield  {author} {\bibinfo {author} {\bibfnamefont {R.}~\bibnamefont
  {Rey}},\ }\href@noop {} {\bibfield  {journal} {\bibinfo  {journal} {J. Phys.
  Chem. B}\ }\textbf {\bibinfo {volume} {112}},\ \bibinfo {pages} {344}
  (\bibinfo {year} {2008})}\BibitemShut {NoStop}%
\bibitem [{\citenamefont {Tkachenko}\ and\ \citenamefont
  {Rabin}(1997)}]{tkachenko_pre}%
  \BibitemOpen
  \bibfield  {author} {\bibinfo {author} {\bibfnamefont {A.~V.}\ \bibnamefont
  {Tkachenko}}\ and\ \bibinfo {author} {\bibfnamefont {Y.}~\bibnamefont
  {Rabin}},\ }\href@noop {} {\bibfield  {journal} {\bibinfo  {journal} {Phys.
  Rev. E}\ }\textbf {\bibinfo {volume} {55}},\ \bibinfo {pages} {778} (\bibinfo
  {year} {1997})}\BibitemShut {NoStop}%
\bibitem [{\citenamefont {Achim}\ \emph {et~al.}(2011)\citenamefont {Achim},
  \citenamefont {Wittkowski},\ and\ \citenamefont
  {L\"{o}wen}}]{Achim_PRE_2011}%
  \BibitemOpen
  \bibfield  {author} {\bibinfo {author} {\bibfnamefont {C.~V.}\ \bibnamefont
  {Achim}}, \bibinfo {author} {\bibfnamefont {R.}~\bibnamefont {Wittkowski}}, \
  and\ \bibinfo {author} {\bibfnamefont {H.}~\bibnamefont {L\"{o}wen}},\
  }\href@noop {} {\bibfield  {journal} {\bibinfo  {journal} {Phys. Rev. E}\
  }\textbf {\bibinfo {volume} {83}},\ \bibinfo {pages} {061712} (\bibinfo
  {year} {2011})}\BibitemShut {NoStop}%
\bibitem [{\citenamefont {Praetorius}\ \emph {et~al.}(2013)\citenamefont
  {Praetorius}, \citenamefont {Voigt}, \citenamefont {Wittkowski},\ and\
  \citenamefont {L\"{o}wen}}]{Simon_PRE_2013}%
  \BibitemOpen
  \bibfield  {author} {\bibinfo {author} {\bibfnamefont {S.}~\bibnamefont
  {Praetorius}}, \bibinfo {author} {\bibfnamefont {A.}~\bibnamefont {Voigt}},
  \bibinfo {author} {\bibfnamefont {R.}~\bibnamefont {Wittkowski}}, \ and\
  \bibinfo {author} {\bibfnamefont {H.}~\bibnamefont {L\"{o}wen}},\ }\href@noop
  {} {\bibfield  {journal} {\bibinfo  {journal} {Phys. Rev. E}\ }\textbf
  {\bibinfo {volume} {87}},\ \bibinfo {pages} {052406} (\bibinfo {year}
  {2013})}\BibitemShut {NoStop}%
\bibitem [{\citenamefont {Rubinstein}\ and\ \citenamefont
  {Glicksman}(1991)}]{glicksman_jcg}%
  \BibitemOpen
  \bibfield  {author} {\bibinfo {author} {\bibfnamefont {E.~R.}\ \bibnamefont
  {Rubinstein}}\ and\ \bibinfo {author} {\bibfnamefont {M.~E.}\ \bibnamefont
  {Glicksman}},\ }\href@noop {} {\bibfield  {journal} {\bibinfo  {journal} {J.
  Cryst. Growth}\ }\textbf {\bibinfo {volume} {112}},\ \bibinfo {pages} {84}
  (\bibinfo {year} {1991})}\BibitemShut {NoStop}%
\bibitem [{\citenamefont {Oswald}\ and\ \citenamefont
  {Pieranski}(2006)}]{oswald_book}%
  \BibitemOpen
  \bibfield  {author} {\bibinfo {author} {\bibfnamefont {P.}~\bibnamefont
  {Oswald}}\ and\ \bibinfo {author} {\bibfnamefont {P.}~\bibnamefont
  {Pieranski}},\ }\href@noop {} {\emph {\bibinfo {title} {Smectic and Columnar
  Liquid Crystals}}}\ (\bibinfo  {publisher} {Taylor and Francis},\ \bibinfo
  {address} {New York},\ \bibinfo {year} {2006})\BibitemShut {NoStop}%
\bibitem [{\citenamefont {Mermin}(1979)}]{Mermin_rmp}%
  \BibitemOpen
  \bibfield  {author} {\bibinfo {author} {\bibfnamefont {N.~D.}\ \bibnamefont
  {Mermin}},\ }\href@noop {} {\bibfield  {journal} {\bibinfo  {journal} {Rev.
  Mod. Phys.}\ }\textbf {\bibinfo {volume} {51}},\ \bibinfo {pages} {591}
  (\bibinfo {year} {1979})}\BibitemShut {NoStop}%
\bibitem [{\citenamefont {Alexander}\ \emph {et~al.}(2012)\citenamefont
  {Alexander}, \citenamefont {g.~Chen}, \citenamefont {Matsumoto},\ and\
  \citenamefont {Kamien}}]{Alexander_rmp}%
  \BibitemOpen
  \bibfield  {author} {\bibinfo {author} {\bibfnamefont {G.~P.}\ \bibnamefont
  {Alexander}}, \bibinfo {author} {\bibfnamefont {B.~G.}\ \bibnamefont
  {g.~Chen}}, \bibinfo {author} {\bibfnamefont {E.~A.}\ \bibnamefont
  {Matsumoto}}, \ and\ \bibinfo {author} {\bibfnamefont {R.~D.}\ \bibnamefont
  {Kamien}},\ }\href@noop {} {\bibfield  {journal} {\bibinfo  {journal} {Rev.
  Mod. Phys.}\ }\textbf {\bibinfo {volume} {84}},\ \bibinfo {pages} {497}
  (\bibinfo {year} {2012})}\BibitemShut {NoStop}%
\bibitem [{\citenamefont {Cremer}\ \emph {et~al.}(2012)\citenamefont {Cremer},
  \citenamefont {Marechal},\ and\ \citenamefont {L\"{o}wen}}]{lowen_epl}%
  \BibitemOpen
  \bibfield  {author} {\bibinfo {author} {\bibfnamefont {P.}~\bibnamefont
  {Cremer}}, \bibinfo {author} {\bibfnamefont {M.}~\bibnamefont {Marechal}}, \
  and\ \bibinfo {author} {\bibfnamefont {H.}~\bibnamefont {L\"{o}wen}},\
  }\href@noop {} {\bibfield  {journal} {\bibinfo  {journal} {Europhys. Lett.}\
  }\textbf {\bibinfo {volume} {99}},\ \bibinfo {pages} {38005} (\bibinfo {year}
  {2012})}\BibitemShut {NoStop}%
\bibitem [{\citenamefont {Wittkowski}\ \emph
  {et~al.}(2011{\natexlab{a}})\citenamefont {Wittkowski}, \citenamefont
  {L\"{o}wen},\ and\ \citenamefont {Brand}}]{RW_prea}%
  \BibitemOpen
  \bibfield  {author} {\bibinfo {author} {\bibfnamefont {R.}~\bibnamefont
  {Wittkowski}}, \bibinfo {author} {\bibfnamefont {H.}~\bibnamefont
  {L\"{o}wen}}, \ and\ \bibinfo {author} {\bibfnamefont {H.~R.}\ \bibnamefont
  {Brand}},\ }\href@noop {} {\bibfield  {journal} {\bibinfo  {journal} {Phys.
  Rev. E}\ }\textbf {\bibinfo {volume} {83}},\ \bibinfo {pages} {061706}
  (\bibinfo {year} {2011}{\natexlab{a}})}\BibitemShut {NoStop}%
\bibitem [{\citenamefont {Wittkowski}\ \emph
  {et~al.}(2011{\natexlab{b}})\citenamefont {Wittkowski}, \citenamefont
  {L\"{o}wen},\ and\ \citenamefont {Brand}}]{RW_preb}%
  \BibitemOpen
  \bibfield  {author} {\bibinfo {author} {\bibfnamefont {R.}~\bibnamefont
  {Wittkowski}}, \bibinfo {author} {\bibfnamefont {H.}~\bibnamefont
  {L\"{o}wen}}, \ and\ \bibinfo {author} {\bibfnamefont {H.~R.}\ \bibnamefont
  {Brand}},\ }\href@noop {} {\bibfield  {journal} {\bibinfo  {journal} {Phys.
  Rev. E}\ }\textbf {\bibinfo {volume} {84}},\ \bibinfo {pages} {041708}
  (\bibinfo {year} {2011}{\natexlab{b}})}\BibitemShut {NoStop}%
\bibitem [{\citenamefont {Backofen}\ and\ \citenamefont
  {Voigt}(2009)}]{Backofen}%
  \BibitemOpen
  \bibfield  {author} {\bibinfo {author} {\bibfnamefont {R.}~\bibnamefont
  {Backofen}}\ and\ \bibinfo {author} {\bibfnamefont {A.}~\bibnamefont
  {Voigt}},\ }\href@noop {} {\bibfield  {journal} {\bibinfo  {journal} {J.
  Phys.: Condens. Matter.}\ }\textbf {\bibinfo {volume} {21}},\ \bibinfo
  {pages} {464109} (\bibinfo {year} {2009})}\BibitemShut {NoStop}%
\bibitem [{\citenamefont {Khare}\ \emph {et~al.}(2003)\citenamefont {Khare},
  \citenamefont {Kodambaka}, \citenamefont {Johnson}, \citenamefont {Petrov},\
  and\ \citenamefont {Greene}}]{Khare_SS}%
  \BibitemOpen
  \bibfield  {author} {\bibinfo {author} {\bibfnamefont {S.~V.}\ \bibnamefont
  {Khare}}, \bibinfo {author} {\bibfnamefont {S.}~\bibnamefont {Kodambaka}},
  \bibinfo {author} {\bibfnamefont {D.~D.}\ \bibnamefont {Johnson}}, \bibinfo
  {author} {\bibfnamefont {I.}~\bibnamefont {Petrov}}, \ and\ \bibinfo {author}
  {\bibfnamefont {J.~E.}\ \bibnamefont {Greene}},\ }\href@noop {} {\bibfield
  {journal} {\bibinfo  {journal} {Surf. Sci.}\ }\textbf {\bibinfo {volume}
  {1-3}},\ \bibinfo {pages} {522} (\bibinfo {year} {2003})}\BibitemShut
  {NoStop}%
\bibitem [{\citenamefont {Lopez-Leon}\ \emph {et~al.}(2012)\citenamefont
  {Lopez-Leon}, \citenamefont {Bates},\ and\ \citenamefont
  {Fernandez-Nieves}}]{Lopez-Leon_pre}%
  \BibitemOpen
  \bibfield  {author} {\bibinfo {author} {\bibfnamefont {T.}~\bibnamefont
  {Lopez-Leon}}, \bibinfo {author} {\bibfnamefont {M.~A.}\ \bibnamefont
  {Bates}}, \ and\ \bibinfo {author} {\bibfnamefont {A.}~\bibnamefont
  {Fernandez-Nieves}},\ }\href@noop {} {\bibfield  {journal} {\bibinfo
  {journal} {Phys. Rev. E}\ }\textbf {\bibinfo {volume} {86}},\ \bibinfo
  {pages} {030702 (R)} (\bibinfo {year} {2012})}\BibitemShut {NoStop}%
\end{thebibliography}%

\end{document}